\documentstyle[aps,eqsecnum,epsf]{revtex}
\begin{document}
\def\I.#1{\it #1}
\def\B.#1{{\bbox#1}}
\def\C.#1{{\cal  #1}}

\title{{\rm DRAFT for IUTAM symposium 1999 \hfill   Version of \today  } \\~~\\
        Anomalous Scaling in Passive Scalar Advection and Lagrangian
        Shape Dynamics}
\author {Itai Arad and  Itamar Procaccia}
\address{Department of~~Chemical Physics, The Weizmann Institute of
         Science, Rehovot 76100, Israel}
\maketitle

%
%

\begin{abstract}
The problem of anomalous scaling in passive scalar advection, especially with
$\delta$-correlated velocity field (the Kraichnan model) has attracted a
lot of
interest since the exponents can be computed analytically in certain limiting
cases. In this paper we focus, rather than on the evaluation of the exponents,
on elucidating the {\em physical mechanism} responsible for the anomaly. We
show
that the anomalous exponents $\zeta_n$ stem from the Lagrangian dynamics of
shapes which characterize configurations of $n$ points in space. Using the
shape-to-shape transition probability, we define an operator whose
eigenvalues
determine the anomalous exponents for all $n$, in all the sectors of the
SO($3$)
symmetry group.
\end{abstract}
\pacs{PACS numbers 47.27.Gs, 47.27.Jv, 05.40.+j}

%
%

\section{Introduction} \label{intro}
In the lecture at the IUTAM symposium the work of our group on the
consequences
of anisotropy on the universal statistics of turbulence has been reviewed.
This
material is available in print, and the interested reader can find it in
\cite{99ALP,98ADKLPS,99ABMP,00KLPS,00ABP,00ALPP}. A short review is
available in
the proceedings of ``Dynamics Days Asia" \cite{00ALP}. In this paper we review
some recent work aimed at understanding the {\em physical mechanism}
responsible
for the anomalous exponents that characterize the statistics of passive
scalars
advected by turbulent velocity fields. We will consider isotropic advecting
velocity fields, but will allow anisotropy in the forcing of the passive
scalar.
In such case the statistical objects like structure functions and correlation
functions are not isotropic. Instead, they are composed of an isotropic and
non-isotropic parts. We overcome this complication by characterizing these
functions in terms of the SO($3$) irreducible representations. Any such
function
can be written as a linear combination of parts which belong to a given
irreducible representation of SO($3$). We will show that each part is
characterized by a set of {\em universal} scaling exponents. The weight of
each
part however will turn out to be non-universal, set by the boundary
conditions.

The SO($3$) classification will appear to be natural once we focus on the
physics of Lagrangian trajectories in the flow. We will see that one can
offer a
satisfactory understanding of the physics of anomalous scaling by
connecting the
the statistics of the passive scalar to the Lagrangian trajectories. This
connection provides a very clear understanding of the physical origin of the
anomalous exponents, relating them to the dynamics in the space of {\em
shapes}
of groups of Lagrangian particles.

%
%

\section{The Kraichnan Model of Passive Scalar Advection}
The model of passive scalar advection with rapidly decorrelating velocity
field
was introduced by R.H. Kraichnan \cite{68Kra} already in 1968. In recent years
\cite{94Kra,94LPF,95GK,95CFKL,96FGLP,96BGK} it was shown to be a fruitful case
model for understanding multiscaling in the statistical description of
turbulent
fields. The basic dynamical equation in this model is for a scalar field
$T(\B.r,t)$ advected by a random velocity field ${\B.u}(\B.r,t)$:
\begin{equation}
\label{advect}
        \big[\partial_t  - \kappa_0 \nabla^2  +
        {\B.u}({\B.r},t) \cdot \bbox{\nabla}\big]
         T({\B.r},t) = f({\B.r},t)\ .
\end{equation}
In this equation $f({\B.r},t)$ is the forcing and $\kappa_0$ is the molecular
diffusivity.  In Kraichnan's model the advecting field ${\B.u}({\B.r},t)$ as
well as the forcing field $f({\B.r},t)$ are taken to be Gaussian, time and
space
homogeneous, and delta-correlated in time:
\begin{equation}
 \left< (u^\alpha(\B.r,t)-u^\alpha(\B.r',t))
         (u^\beta(\B.r,t')-u^\beta(\B.r',t'))
\right>_{\B.u} = h^{\alpha\beta}(\B.r-\B.r') \delta(t-t') \ ,
\end{equation}
where the ``eddy-diffusivity'' tensor $h^{\alpha\beta}(\B.r)$ is defined by
\begin{equation}
h^{\alpha\beta}(\B.r) =
    \left({r \over \Lambda}\right)^\xi
    (\delta^{\alpha\beta}-{\xi \over d-1+\xi}{r^\alpha r^\beta \over r^2})\ ,
    \quad \eta \ll r \ll \Lambda \ .
\label{eddy}
\end{equation}
Here $\eta$ and $\Lambda$ are the inner and outer scale for the velocity
fields,
and the coefficients are chosen such that $\partial_\alpha h^{\alpha\beta}=0$.
The averaging $\langle \dots \rangle_{\B.u}$ is done with respect to the
realizations of the velocity field.

The forcing $f$ is also taken white in time and Gaussian:
\begin{equation}
 \left< f(\B.r,t)f(\B.r',t')\right>_{f} = \Xi(\B.r-\B.r') \delta(t-t') \ .
\end{equation}
Here the average is done with respect to realizations of the forcing. The
forcing is taken to act only on the large scales, of the order of $L$ (with a
compact support in Fourier space). This means that the function $\Xi(r)$ is
nearly constant for $r \ll L$ but is decaying rapidly for $r>L$.

>From the point of view of the statistical theory one is interested mostly
in the
scaling exponents characterizing the structure functions
\begin{equation}
S_{2n}(\B.r_1,\B.r_2)= \left<
(T(\B.r_1,t)-T(\B.r_2,t))^{2n}\right>_{\B.u,f} \ .
\label{S2n}
\end{equation}
For isotropic forcing one expects $S_{2n}$ to depend only on the distance
$R \equiv |\B.r_1-\B.r_2|$ such that in the scaling regime
\begin{equation}
    S_{2n}(R) \propto R^{\zeta_{2n}} \propto [S_2(R)]^{n}
\left({L \over R} \right)^{\delta_{2n}}\ .
\end{equation}
In this equation we introduced the ``normal'' $(n\zeta_2)$ and the anomalous
$(\delta_{2n})$ parts of the scaling exponents
$\zeta_{2n} = n \zeta_2-\delta_{2n}$. The first part can be obtained from
dimensional considerations, but the anomalous part cannot be guessed from
simple
arguments.

When the forcing is anisotropic, the structure functions depend on the vector
distance $\B.R=\B.r_1-\B.r_2$. In this case we can represent them in terms of
spherical harmonics,
\begin{equation}
    S_{2n}(\B.R) =\sum_{\ell,m} a_{\ell,m}(R)Y_{\ell,m}(\hat \B.R) \ ,
\end{equation}
where $\hat \B.R\equiv \B.R/R$. This is a case in which the statistical object
is a scalar function of one vector, and the appropriate irreducible
representation of the SO($3$) symmetry group are obvious. We are going to
explain in the next section that the coefficients $a_{\ell,m}(R)$ are expected
to scale with a universal leading scaling exponent $\zeta_{2n}^{(\ell)}$. The
exponent will turn out to be $\ell$ dependent but not $m$ dependent.

Theoretically it is natural to consider correlation functions rather than
structure functions. The $2n$-order correlation functions are defined as
\begin{equation}
    F_{2n}(\B.r_1,\ldots,\,\B.r_{2n}) \equiv
        \left< T(\B.r_1)T(\B.r_2) \dots T(\B.r_{2n}) \right>_{\B.u,f} \ .
\label{F2n}
\end{equation}
For separations $r_{ij}\to 0$ the correlation functions converges to
$\left< T^{2n} \right>_{\B.f}$, whereas for $r_{ij}\to L$ decorrelation
leads to convergence to $\left< T \right>_{\B.u,f}^n$. For all
$r_{ij}\approx O(r)\ll L$ one expects a behaviour according to
\begin{equation}
    F_{2n}(\B.r_1, \ldots,\,\B.r_{2n})= L^{n(2-\xi)}(c_0 + \cdots +
        c_k(r/L)^{\zeta_{2n}} \tilde F_{2n}(\tilde \B.r_1, \dots ,
        \tilde \B.r_{2n})+\cdots) \ ,
\end{equation}
where $\tilde F_{2n}$ is a scaling function depending on $\tilde\B.r_i$ which
denote a set of dimensionless coordinates describing the configuration of the
$2n$ points. The exponents and scaling functions are expected to be universal,
but not the $c$ coefficients, which depend on the details of forcing.

It has been shown \cite{95GK} that the anomalous exponents $\zeta_{2n}$ can be
obtained by solving for the zero modes of the exact differential equations
which
are satisfied by $F_{2n}$. The equations for the zero modes read
\begin{equation}
\big[- \kappa \sum_{i} \nabla^2_i + \hat {\cal B}_{2n}\big]
    {\cal F}_{2n}({\bf r}_1,{\bf r}_2,...,{\bf r}_{2n}) = 0 \ .
\label{difeq}
\end{equation}
The operator $\hat {\cal B}_{2n}\equiv \sum_{i>j}^{2n} \hat{\cal B} _{ij}$,
and
$\hat{\cal B} _{ij}$ are defined by
\begin{equation}
    \hat{\cal B}_{ij}\equiv \hat {\cal B}(\B.r_i, \B.r_j) =
    h^{\alpha\beta}(\B.r_i - \B.r_j) \partial^2 /
    \partial r^\alpha_i \partial r^\beta_j  \ .
\end{equation}

%
%

\section{Lagrangian Trajectories, Correlation Functions and Shape Dynamics}

An elegant approach to the correlation functions is furnished by Lagrangian
dynamics \cite{00ALP,94SG,GZ,98GPZ,VF}. In this formalism one recognizes that
the actual value of the scalar at position $\B.r$ at time $t$ is determined by
the action of the forcing along the Lagrangian trajectory from $t=-\infty$ to
$t$:
\begin{equation}
  T(\B.r_0,t_0)=\int_{-\infty}^{t_0} dt
    \left< f(\B.r(t),t) \right>_{\B.\eta} \ ,
\label{path}
\end{equation}
with the trajectory $\B.r(t)$ obeying
\begin{eqnarray}
    \B.r(t_0) &=& \B.r_0  \ , \nonumber \\
    \partial_t \B.r(t) &=& \B.u(\B.r(t),t) +
        \sqrt{2 \kappa}{\B.\eta}(t) \ ,
\label{traject1}
\end{eqnarray}
and $\B.\eta$ is a vector of  zero-mean independent Gaussian white random
variables,
$\left< \eta^\alpha (t) \eta^\beta (t') \right> = \delta^{\alpha\beta}
\delta(t-t')$. With this in mind, we can rewrite $F_{2n}$ by substituting each
factor of $T(\B.r_i)$  by its representation (\ref{path}). Performing the
averages  over the random forces, we end up with
\begin{eqnarray}
    F_{2n}(\B.r_1, \ldots, \B.r_{2n},t_0) =
        \Big< \int_{-\infty}^{t_0} dt_1 \cdots dt_n
        \Big[ \Xi(\B.r_1(t_1)-\B.r_2(t_1)) \cdots \\
        \times \Xi(\B.r_{2n-1}(t_n)-\B.r_{2n}(t_n))
        + \hbox{permutations} \Big] \Big>_{\B.u,\{\B.\eta_i\}} \ ,
\label{FnXi}
\end{eqnarray}
To understand the averaging procedure recall that each of the trajectories
$\B.r_i$ obeys an equation of the form (\ref{traject1}), where $\B.u$ as well
as $\{\B.\eta_i\}_{i=1}^{2n}$ are independent stochastic variables whose
correlations are given above. Alternatively, we refer the reader to section
II of \cite{VF}, where the above analysis is carried out in detail.

In considering Lagrangian trajectories of {\em groups} of particles, we should
note that every initial configuration is characterized by a {\em center of
mass}, say $\B.R$, a {\em scale} $s$ (say the radius of gyration of the
cluster
of particles) and a {\em shape} $\B.Z$. In ``shape" we mean here all the
degrees
of freedom other than the scale and $\B.R$: as many angles as are needed to
fully determine a shape, in addition to the Euler angles that fix the
shape orientation with respect to a chosen frame of coordinates. Thus a
group of
$2n$ positions $\{\B.r_i\}$ will be sometime denoted below as
$\{\B.R,s,\B.Z\}$.

One component in the evolution of an initial configuration is a rescaling
of all
the distances which increase on the average like $t^{1/\zeta_2}$; this
rescaling
is analogous to Richardson diffusion. The exponent $\zeta_2$ which determines
the scale increase is also the characteristic exponent of the second order
structure function \cite{68Kra}. This has been related to the exponent
$\xi$ of
(\ref{eddy}) according to $\zeta_2=2-\xi$. After factoring  out this overall
expansion we are left with a normalized `shape'. It is the evolution of this
shape that determines the anomalous exponents.

Consider a final shape ${\B.Z}_{0}$  with an overall scale $s_0$ which is
realized at $t=0$. This shape has evolved during negative times. We fix a
scale $s>s_0$ and examine the shape when the configuration reaches the scale
$s$ for the last time before reaching the scale $s_0$. Since the trajectories
are random the shape $\B.Z$ which is realized at this time is taken from a
distribution $\gamma(\B.Z;\B.Z_0,s\to s_0)$. As long as the advecting velocity
field is scale invariant, this distribution can depend only on the ratio
$s/s_0$. \\
Next, we use the shape-to-shape transition probability to define an operator
$\hat \gamma (s/ s_0)$ on the space of functions $\Psi(\B.Z)$ according to
\begin{equation}
    [\hat \B.\gamma (s/ s_0) \Psi](Z_0) =
        \int d\B.Z \gamma(\B.Z;\B.Z_0,s \to s_0) \Psi(\B.Z)
\end{equation}
We will be interested in the eigenfunction and eigenvalues of this operator.
This operator has two important properties. First, for an isotropic statistics
of the velocity field the operator is isotropic. This means that this operator
commutes with all rotation operators on the space of functions
$\Psi(\B.Z)$. In
other words, if ${\cal O}_\Lambda$ is the rotation operator that takes the
function $\Psi(\B.Z)$ to the new function $\Psi(\Lambda^{-1} \B.Z)$, then
\begin{equation}
    {\cal O}_\Lambda \hat\B.\gamma =\hat\B.\gamma {\cal O}_\Lambda \ .
\label{rot-lam}
\end{equation}
This property follows from the obvious symmetry of the Kernel
$\gamma(\B.Z;\B.Z_0,s\to s_0)$ to rotating $\B.Z$ and $\B.Z_0$ simultaneously.
Accordingly the eigenfunctions of $\hat \B.\gamma$ can be classified according
to the irreducible representations of SO($3$) symmetry group. We will denote
these eigenfunctions as $B_{q\ell m}(\B.Z)$. Here $\ell=0,1,2,\dots$,
$m=-\ell, -\ell+1,\dots \ell$ and $q$ stands for a running index if there is
more than one representation with the same $\ell,m$. The fact that the
$B_{q\ell m}(\B.Z)$ are classified according to the irreducible
representations
of SO($3$) in manifested in the action of the rotation operators upon them:
\begin{equation}
    {\cal O}_\Lambda B_{q\ell m} =
    \sum_{m'} D^{(\ell)}_{m'm}(\Lambda) B_{q\ell m'}
\label{rot-operator}
\end{equation}
where $D^{(\ell)}_{m'm}(\Lambda)$ is the SO($3$) $\ell \times \ell$
irreducible
matrix representation.

The second important property of $\hat \B.\gamma$ follows from the
$\delta$-correlation in time of the velocity field. Physically this means that
the future trajectories of $n$ particles are statistically independent of
their
trajectories in the past. Mathematically, it implies for the kernel that
\begin{equation}
  \gamma(\B.Z;\B.Z_0,s\to s_0)  = \int d\B.Z_1 \gamma(\B.Z;\B.Z_1,s\to s_1)
    \gamma(\B.Z_1;\B.Z_0,s_1\to s_0) \ , \quad s>s_1>s_0
\end{equation}
and in turn, for the operator, that
\begin{equation}
\hat \B.\gamma (s/ s_0) =\hat \B.\gamma (s/ s_1)\hat \B.\gamma (s_1/ s_0) \ .
\end{equation}

Accordingly, by a successive application of $\hat\B.\gamma(s/s_0)$ to an
arbitrary eigenfunction, we get that the eigenvalues of $\hat\B.\gamma$
have to
be of the form $\alpha_{q,\ell}=(s/s_0)^{\zeta_{2n}^{(q,\ell)}}$:
\begin{equation}
  ({s \over s_0})^{\zeta_{2n}^{(q,\ell)}} B_{q\ell m}(\B.Z_0) =
    \int d \B.Z\gamma(\B.Z; \B.Z_0, s \to s_0) B_{q\ell m}(\B.Z)
\label{eig}
\end{equation}
Notice that the eigenvalues are not a function of $m$. This follows from
Schur's
lemmas \cite{62Ham}, but can be also explained from the fact that the rotation
operator mixes the different $m$'s (\ref{rot-operator}): Take an eigenfunction
$B_{q \ell m}(\B.Z)$, and act on it once with the operator
${\cal O}_\Lambda \hat\B.\gamma(s/s_0)$ and once with the operator
$\hat\B.\gamma(s/s_0){\cal O}_\Lambda$. By virtue of (\ref{rot-lam}) we should
get that same result, but this is only possible if all the eigenfunctions with
the same $\ell$ {\em and} the same $q$ share the same eigenvalue.

To proceed we want to introduce into the averaging process in (\ref{FnXi}) by
averaging over Lagrangian trajectories of the $2n$ particles. This will
allow us
to connect the shape dynamics to the statistical objects. To this aim consider
any set of Lagrangian trajectories that started at $t=-\infty$ and end up at
time $t=0$ in a configuration characterized by a scale $s_0$ and center of
mass
$\B.R_0=0$. A full measure of these have evolved through the scale $L$ or
larger. Accordingly they must have passed, during their evolution from time
$t=-\infty$ through a configuration of scale $s>s_0$ at least once. Denote now
\begin{equation}
    \mu_{2n}(t,R,\B.Z;s\to s_0,\B.Z_0)dtd\B.Rd\B.Z
\nonumber
\end{equation}
as the probability that this set of $2n$ trajectories crossed the scale $s$
for
the last time before reaching $s_0,\B.Z_0$, between $t$ and $t+dt$, with a
center of mass between $\B.R$ and $\B.R+d\B.R$ and with a shape between $\B.Z$
and $\B.Z+d\B.Z$.

In terms of this probability we can rewrite Eq.(\ref{FnXi}) (displaying, for
clarity, $\B.R_0=0$ and $t=0$) as
\begin{eqnarray}
 && F_{2n}(\B.R_0=0,s_0,\B.Z_0,t=0) = \int d\B.Z\int_{-\infty}^0 dt \int d\B.R
    \mu_{2n}(t,R,\B.Z; s \to s_0, \B.Z_0) \nonumber \\
 && \times\left< \int_{-\infty}^{0}dt_1\cdots dt_n
    \Big[ \Xi(\B.r_1(t_1)-\B.r_2(t_1)) \cdots
        \Xi(\B.r_{2n-1}(t_n) - \B.r_{2n}(t_n)) +\hbox{perms}\Big]
    \Big \vert (s; \B.R, \B.Z, t) \right> _{\B.u,\B.\eta_i} \ .
\label{Fntraj}
\end{eqnarray}
The meaning of the conditional averaging is an averaging over all the
realizations of the velocity field and the random $\B.\eta_i$ for which
Lagrangian trajectories that ended up at time $t=0$ in $\B.R=0,s_0, \B.Z_0$
passed through $\B.R ,s,\B.Z$ at time $t$.

Next, the time integrations in Eq.(\ref{Fntraj}) are split to the interval
$[-\infty,t]$ and $[t,0]$ giving rise to $2^n$ different contributions:
\begin{equation}
    \int_{-\infty}^t dt_1 \cdots \int_{-\infty}^t dt_n +
    \int_{t}^0 dt_1 \int_{-\infty}^{t} dt_2 \cdots \int_{-\infty}^{t} dt_n
    + \dots
\end{equation}
Consider first the contribution with $n$ integrals in the domain $[-\infty,t]$.
It follows from the delta-correlation in time of the velocity field, that
we can
write
\begin{eqnarray}
   && \left< \int_{-\infty}^t dt_1\cdots dt_n
    \Big[ \Xi(\B.r_1(t_1)-\B.r_2(t_1)) \cdots
          \Xi(\B.r_{2n-1}(t_n)-\B.r_{2n}(t_n)) + \hbox{perms} \Big]
    \Big \vert (s; \B.R, \B.Z,t) \right> _{\B.u,\B.\eta_i} \nonumber \\
   && = \left< \int_{-\infty}^t dt_1 \cdots dt_n
    \Big[\Xi(\B.r_1(t_1)-\B.r_2(t_1)) \cdots \Xi(\B.r_{2n-1}(t_n) -
           \B.r_{2n}(t_n)) + \hbox{perms} \Big]
   \right> _{\B.u,\B.\eta_i} \nonumber \\
   && = F_{2n}(\B.R, s, \B.Z, t)=F_{2n}(s, \B.Z) \ .
\end{eqnarray}
The last equality follows from translational invariance in space-time.
Accordingly the contribution with $n$ integrals in the domain
$[-\infty,t]$
can be written as
\begin{equation}
    \int d\B.Z F_{2n}(s,\B.Z) \int_{-\infty}^0 dt \int d\B.R~~
    \mu_{2n}(t,R,\B.Z;s\to s_0,\B.Z_0) \ .
\end{equation}
We identify the shape-to-shape transition probability:
\begin{equation}
    \gamma(\B.Z;\B.Z_0,s \to s_0)=\int_{-\infty}^0 dt\int d\B.R~~
    \mu_{2n}(t,R,\B.Z;s \to s_0, \B.Z_0) \ .
\end{equation}

Finally, putting all this added wisdom back in Eq.(\ref{Fntraj}) we end up with
\begin{equation}
    F_{2n}(s_0, \B.Z_0) = I + \int d\B.Z \gamma(\B.Z; \B.Z_0, s \to s_0)
        F_{2n}(s,\B.Z)\ . \label{crucial}
\label{split}
\end{equation}
Here $I$ represents all the contributions with one or more time integrals
in the
domain $[t,0]$.  The key point now is that only the term with $n$ integrals in
the domain $[-\infty,t]$ contains information about the evolution of $2n$
Lagrangian trajectories that probed the forcing scale $L$. Accordingly, the
term
denoted by $I$ cannot contain information about the leading anomalous scaling
exponent belonging to $F_{2n}$, but only of lower order exponents. The
anomalous
scaling dependence of the LHS of Eq.(\ref{crucial}) has to cancel against the
integral containing $F_{2n}$ without the intervention of $I$.

Representing now
\begin{eqnarray}
  F_{2n}(s_0,\B.Z_0) &=& \sum_{q\ell m} a_{q,\ell m}(s_0)
    B_{q\ell m}(\B.Z_0)\ ,\nonumber \\
  F_{2n}(s,\B.Z) &=& \sum_{q\ell m} a_{q, \ell m}(s)
    B_{q\ell m}(\B.Z) \ ,\nonumber \\
I&=& \sum_{q\ell m} I_{q \ell m} B_{q\ell m}(\B.Z_0)
\end{eqnarray}
and substituting on both sides of Eq.(\ref{crucial}) and using Eq.(\ref{eig})
we find, due to the linear independence of the eigenfunctions $B_{q\ell m}$
\begin{equation}
  a_{q, \ell m}(s_0) = I_{q \ell m} +
  \left(\frac{s}{s_0}\right)^{\zeta_{2n}^{(q,\ell)}} a_{q, \ell m}(s)
\end{equation}
To leading order the contribution of $I_{q \ell m}$ is neglected, leading
to the
conclusion that {\em the spectrum of anomalous exponents of the correlation
functions is determined by the eigenvalues of the shape-to-shape transition
probability operator}. Calculations show that the leading exponent in the
isotropic sector is always smaller than the leading exponents in all other
sectors. This gap between the leading exponent in the isotropic sector to the
rest of the exponents determines the rate of decay of anisotropy upon
decreasing
the scale of observation.

%
%

\section{Concluding remarks}
The derivation presented above has used explicitly the properties of the
advecting field, in particular the $\delta$-correlation in time.
Accordingly it
cannot be immediately generalized to more generic situations in which there
exist time correlations. Nevertheless we find it pleasing that at least in the
present case we can trace the physical origin of the exponents anomaly,
and connect it to the underlying dynamics. In more generic cases the
mechanisms
may be more complicated, but one should still keep the lesson in mind - higher
order correlation functions depend on many coordinates, and these define a
configuration in space. The scaling properties of such functions may very well
depend on how such configurations are reached by the dynamics. Focusing on
static objects like structure functions of one variable may be insufficient
for
the understanding of the physics of anomalous scaling.

%
%

\acknowledgments
This work has been supported in part by the German-Israeli Foundation,  The
European Commission under the TMR program and the Naftali and Anna
Backenroth-Bronicki Fund for Research in Chaos and Complexity.

%
%

\end{document}